\documentclass[12pt]{article}

\usepackage{latexsym}
\usepackage{comment}
\usepackage[font=small,labelfont=bf]{caption}
\usepackage[numbers,sort&compress]{natbib}
\usepackage{graphics}
\usepackage{graphicx}
\usepackage{epstopdf}
\usepackage{amssymb}
\usepackage{amsmath}
\usepackage{tabularx}
\usepackage{caption}
\usepackage{subcaption}
\usepackage{hyperref}
\usepackage{tabularx}
\usepackage{gensymb}
\usepackage{tensor}
\usepackage{mathtools,amsmath,amssymb,mathrsfs,color,esint}
\usepackage{array}
\usepackage{makecell}
\usepackage[normalem]{ulem}
\usepackage{float}
\usepackage{dcolumn}

\begin{document}

\title{Veiled Singularities in Einstein--Weyl Gravity:\\
Stability and Physical Interpretation of Horizonless Solutions}

\author{A. M.~Bonanno${}^{1,2}$, S.~Silveravalle${}^{3,4,5}$ and A.~Spina${}^{2,6}$ \\
{\footnotesize${}^{1}$ INAF, Osservatorio Astrofisico di Catania, via S. Sofia 78, 95123 Catania, Italy  }\\[-0.15cm]
{ \footnotesize${}^{2}$    INFN, Sezione di Catania,  via S. Sofia 64, 95123,Catania, Italy.} \\[-0.15cm]
{\footnotesize${}^{3}$ SISSA - International School for Advanced Studies, Via Bonomea 265, 34136 Trieste, Italy}\\[-0.15cm]
{\footnotesize${}^{4}$ INFN, Sezione di Trieste, Via Valerio 2, 34127 Trieste, Italy} \\[-0.15cm]
{\footnotesize${}^{5}$  IFPU - Institute for Fundamental Physics of the Universe, Via Beirut 2, 34151 Trieste, Italy}\\[-0.15cm]
{\footnotesize${}^{6}$ Department of Physics and Astronomy, Universit\`a di Catania, via S. Sofia 64, 95123 Catania, Italy}\\[0.2cm]
{\itshape\small Dedicated to the memory of Kellogg Stelle (1948--2025),}\\[-0.15cm]
{\itshape\small whose pioneering work on quadratic gravity continues to inspire us.}}
\date{}
\maketitle

\begin{abstract}

We investigate a class of horizonless solutions in Einstein--Weyl gravity, corresponding to the so-called attractive naked singularities of the $(-2,2)$ type. In contrast to General Relativity, where naked singularities are generically unstable and excluded by the cosmic censorship conjecture, we show that these configurations are linearly stable under tensor perturbations. By numerically evolving the perturbation equations in the time domain, we find that all modes decay with characteristic oscillatory tails, a behavior consistent with the dynamics of massive field perturbations in quadratic gravity. This establishes that attractive naked singularities in Einstein--Weyl gravity are dynamically stable and can persist as stationary configurations. 
We argue that these horizonless configurations are observationally concealed, and therefore we refer to them as \emph{veiled singularities}. Their stability and phenomenological similarity to black holes suggest that they may represent viable horizonless alternatives in higher-derivative theories of gravity, offering a novel perspective on the interplay between singularity resolution, stability, and effective field dynamics beyond Einstein’s theory.
\end{abstract}

\section{Introduction}

One of the most common arguments invoked to justify the enormous efforts devoted to the construction of a consistent and complete theory of quantum gravity is the expectation that such a theory should provide a clear and well-defined physical mechanism capable of removing the singularities that inevitably arise in the framework of classical General Relativity~\cite{Penrose:1964wq,Hawking:1970zqf,Hawking:1973uf}. These singularities, where curvature invariants diverge and the predictive power of the theory breaks down, are generally interpreted as signals of the incompleteness of Einstein’s equations at very high energies or in regions of extreme curvature, such as the cores of black holes or the initial Big Bang singularity. 

From this perspective, the problem of singularity resolution is not only of technical importance but is also deeply connected with the conceptual foundations of gravitational physics. In particular, understanding how quantum effects can regularize or avoid the formation of classical singularities would offer a crucial step toward reconciling General Relativity with the principles of quantum mechanics. Such a mechanism would also have profound implications for the long-standing information-loss problem in black hole physics~\cite{Hawking:1976ra,Harlow:2014yka,Polchinski:2016hrw}, since it would invalidate the classical argument that information carried by infalling matter is unavoidably destroyed when it encounters the curvature singularity. Instead, a consistent quantum theory of gravity capable of removing the singularity would allow for the preservation of quantum information, providing a natural and physically appealing resolution to this paradox.

Despite the fundamental importance of this issue, first-principle investigations aimed at uncovering the mechanisms that might lead to singularity resolution remain relatively scarce~\cite{UV-grav1,UV-grav2,UV-grav3,UV-grav4,Bosma:2019aiu,Borissova:2023kzq}. Historically, the most widely accepted and tractable approach has relied on incorporating quantum effects in a semiclassical framework, where the backreaction of quantum fields on the spacetime geometry is taken into account through the vacuum polarization terms in the effective field equations~\cite{Poisson:1988wc,Arrechea:2021xkp,Arrechea:2023wgy,Arrechea:2023oax}. Although this semiclassical treatment does not provide a full quantization of gravity, it has proven extremely valuable in revealing the qualitative impact of quantum corrections on classical gravitational dynamics, and in particular in suggesting how singularities might be softened or even removed once these corrections are included.

Motivated by these considerations, considerable attention has recently been devoted to the study of classical gravity theories augmented by quadratic curvature corrections, which naturally arise as the leading higher-order terms in the effective action of several quantum gravity scenarios. The inclusion of quadratic curvature invariants represents the simplest and most theoretically consistent way to extend Einstein’s theory while preserving general covariance and locality, and it captures the first quantum corrections expected from the renormalization of the gravitational action~\cite{tHooft:1974toh,Benedetti:2013jk,Zwiebach:1985uq}.

The most general parity-even, diffeomorphism-invariant quadratic action in four spacetime dimensions can be written as
\begin{equation}\label{action}
\mathcal{S} = \int \mathrm{d}^4x \, \sqrt{-g} 
\left[
  \gamma R 
  - \alpha\, C^{\mu\nu\rho\sigma} C_{\mu\nu\rho\sigma}
  + \beta\, R^2 
  + \eta\, \mathcal{G}
\right],
\end{equation}
where $\mathcal{G}$ denotes the topological Gauss--Bonnet term, $R$ is the Ricci scalar, $C_{\mu\nu\rho\sigma}$ is the Weyl tensor, $\gamma = 1/(16\pi G)$, and $\alpha$, $\beta$, and $\eta$ are dimensionless coupling constants. The $\mathcal{G}$ term, being a total derivative in four dimensions, does not contribute to the local equations of motion but can play a role in topologically nontrivial spacetimes. 

When interpreted as a fundamental quantum field theory of gravity, the action~\eqref{action} is well known to be renormalizable~\cite{Stelle:1976gc}, a remarkable property that sets it apart from pure Einstein gravity. However, renormalizability comes at the price of unitarity violation, which manifests itself through the presence of a massive spin-2 excitation with a wrong-sign kinetic term in the propagator. This additional degree of freedom is associated with an Ostrogradsky ghost and is usually regarded as a signal of the perturbative inconsistency or incompleteness of the theory.

A detailed perturbative analysis of the quadratic theory reveals three distinct dynamical modes: the familiar massless graviton of General Relativity, a massive scalar mode with mass $m_0 = \sqrt{\gamma / 6\beta}$ arising from the $R^2$ term, and a massive spin-2 mode with mass $m_2 = \sqrt{\gamma / 2\alpha}$ originating from the Weyl-squared term~\cite{Stelle:1977ry}. While the presence of the latter mode is problematic from a quantum-field-theoretic standpoint, its physical interpretation at the classical level is far subtler. In fact, even the simpler Einstein--Weyl sector of the theory (obtained by setting $\beta = 0$) exhibits a remarkably rich structure of classical solutions, including nontrivial black holes, wormholes, and regular geometries that do not have counterparts in General Relativity~\cite{Lu:2015cqa,Lu:2015psa,perkinsthesis,Goldstein:2017rxn,Bonanno:2019rsq,Podolsky:2019gro,Podolsky:2018pfe,Silveravalle:2022wij,Bonanno:2022ibv,Daas:2022iid,Silveravalle:2023lnl}.

This raises the important question of whether some of these new solutions could be classically stable and physically relevant. If stable, they could provide consistent effective descriptions of gravitational phenomena beyond the Einstein regime, without requiring the theory to be regarded as fundamental in the quantum sense. In this view, the quadratic action~\eqref{action} may be interpreted as an effective low-energy truncation of a more complete underlying theory, capable of capturing essential features of gravitational dynamics in regimes where higher-curvature corrections become significant.

In the present work, we focus our attention on a remarkable class of solutions arising within the framework of Einstein--Weyl gravity that have no counterparts in classical General Relativity: the so-called \emph{attractive naked singularities} of the $(-2,2)$ type. These configurations, first systematically analyzed by Holdom and collaborators~\cite{Holdom:2002xy,Holdom:2016nek,Holdom:2019bdv,Holdom:2020onl,Holdom:2022zzo,Holdom:2022fsm}, have attracted considerable interest as potential \emph{black-hole mimickers}. Their external geometry closely resembles that of the Schwarzschild solution, exhibiting a strong gravitational potential and characteristic light deflection, yet crucially, they lack an event horizon. As a consequence, they represent globally naked configurations, where the central singularity remains visible to distant observers. From a phenomenological standpoint, such objects could in principle reproduce several observable signatures commonly attributed to black holes while avoiding the causal disconnection associated with horizon formation.

In the context of classical General Relativity, the occurrence of naked singularities has long been regarded as a pathological feature, in direct tension with the cosmic censorship conjecture, which asserts that physically reasonable matter configurations cannot give rise to singularities visible to distant observers. This conjecture, though never rigorously proven in full generality, has been supported by several analytical and numerical results. A particularly notable example is the series of works by Christodoulou~\cite{Christodoulou:1984djm,Christodoulou:1991yfa,Christodoulou:1994hg,Christodoulou:1999}, who demonstrated that although naked singularities can indeed form in the gravitational collapse of a massless scalar field, such configurations are non-generic: they are unstable under infinitesimal perturbations of the initial data and therefore do not persist in realistic dynamical evolutions. This result, among others, has strongly reinforced the prevailing view that within Einstein’s theory, naked singularities are dynamically excluded and should not appear in physically relevant gravitational processes.

The situation, however, changes dramatically once higher-curvature corrections are included.  
In particular, we shall perform a detailed analysis of the linear stability of these $(-2,2)$ configurations under small perturbations, and contrast their behavior with the well-established instability of naked singularities in General Relativity. One of the most intriguing outcomes of our analysis is that the $(-2,2)$ family of attractive naked singularities, despite the divergence in curvature invariants at the center, turns out to be linearly stable under tensor perturbations. This result further indicates that the inclusion of higher-derivative terms fundamentally alters the dynamical landscape of General Relativity: configurations that are forbidden or unstable in Einstein’s theory may become dynamically consistent once quadratic corrections are taken into account.

The existence of linearly stable naked singularities within Einstein--Weyl gravity thus has profound implications. It suggests that the boundary between black holes and horizonless compact objects may be less clear-cut than previously thought, and that new gravitational phases could emerge in regimes where higher-curvature effects become significant. From a theoretical perspective, these solutions provide a laboratory for exploring the interplay between stability, causality, and higher-derivative dynamics, while from an observational standpoint,  they could offer a viable alternative to classical black holes, potentially distinguishable through precise measurements quasi-normal modes, or accretion-disk signatures.

\section{Spherically symmetric solutions in Einstein-Weyl gravity}

The Einstein-Weyl theory is defined by the action 
\begin{equation}\label{Action}
S = \int \mathrm{d}^4 x \sqrt{-g} \left[\gamma R -\alpha C_{\mu \nu \rho \sigma}C^{\mu \nu \rho \sigma}\right],
\end{equation}
where $\gamma=1/16\pi G$, $C_{\mu\nu\rho\sigma}$ is the Weyl tensor, and $\alpha$ is a dimensionless parameter. This theory includes all solutions of the full quadratic gravity theory in the sector with vanishing Ricci scalar. In particular, every static and asymptotically flat black hole solution in quadratic gravity also satisfies the equations of the Einstein-Weyl theory~\cite{Bonanno:2019rsq,Lu:2015cqa,Lu:2015psa}. 
The equations of motion are obtained by minimizing the action~\eqref{Action}, yielding
\begin{equation}\label{eomham}
\mathcal{H}_{\mu\nu}=\gamma\left(R_{\mu\nu}-\frac{1}{2}R\,g_{\mu\nu}\right)-4\alpha\left(\nabla^\rho\nabla^\sigma+\frac{1}{2}R^{\rho\sigma}\right)C_{\mu\rho\nu\sigma}=0.
\end{equation}
Due to the tracelessness of the Weyl tensor, taking the trace of~\eqref{eomham} gives
\begin{equation}\label{trace}
    \tensor{\mathcal{H}}{^\mu_\mu}=\gamma R=0.
\end{equation}
To study static, spherically symmetric spacetimes, we adopt the metric ansatz
\begin{equation}\label{metric}
\mathrm{d}s^2=-h(r)\mathrm{d}t^2+\frac{\mathrm{d}r^2}{f(r)}+r^2\mathrm{d}\Omega^2.
\end{equation}
While the equations~\eqref{eomham} are third order in $h(r)$ and $f(r)$, using~\eqref{trace} allows one to reduce the system to two second order ordinary differential equations~\cite{ Lu:2015psa,Bonanno:2022ibv}:
\begin{equation}\label{eom}
\begin{split}
&4 h(r)^2 \big(r f'(r)+f(r)-1\big)+r h(r) \big(r f'(r) h'(r)
+2 f(r) \big(r h''(r)+2 h'(r)\big)\big)\\
&-r^2 f(r) h'(r)^2=0,\\[0.3cm]
&\alpha  r^2 f(r) h(r) \big(r f'(r)+3 f(r)\big) h'(r)^2+2 r^2 f(r) h(r)^2 h'(r) \big(\alpha  r f''(r)+\alpha  f'(r)-\gamma  r\big)\\
&+h(r)^3 \big(r \big(3 \alpha  r f'(r)^2-4 \alpha  f'(r)+2
   \gamma  r\big)-2 f(r) \big(4 \alpha +2 \alpha  r^2 f''(r)-2 \alpha  r f'(r)+\gamma  r^2\big)\\
&+8 \alpha  f(r)^2\big)-\alpha  r^3 f(r)^2 h'(r)^3=0.
\end{split}
\end{equation}
Since these equations cannot be solved analytically, we must resort to numerical methods and approximations to extract meaningful results. At large distances, we impose asymptotic flatness. Following the approach in~\cite{Lu:2015psa, Stelle:1977ry}, we expand the metric functions as
\begin{equation}
h(r)=1+\epsilon\,V(r),\qquad f(r)=1+\epsilon\,W(r)
\end{equation}
and linearize equations~\eqref{eom} in $\epsilon$. Requiring that $h(r),f(r)\to 1$ as $r\to\infty$, we obtain the weak-field solution~\cite{Bonanno:2022ibv, Bonanno:2021zoy}
\begin{equation}\begin{split} \label{eq|weakfield}
h(r)=\, &1-\frac{2\,M}{r}+2S^-_2 \frac{e^{-m_2\, r}}{r}, \\
f(r)=\, &1-\frac{2\,M}{r}+S^-_2 \frac{e^{-m_2\, r}}{r}(1+m_2\, r),
\end{split}\end{equation}
where $m_2^2=\frac{\gamma}{2\alpha}$ is the mass of the spin-2 particle, and $M$ is $G$ times the ADM mass. These solutions are exponentially suppressed corrections to the Schwarzschild metric; in particular, the correction to the gravitational potential in the Newtonian limit $\phi(r)=\frac{1}{2}(h(r)-1)$ takes precisely the form of a Yukawa potential, as expected from the presence of a massive mediator. For this reason, we define the parameter $S_2^-$ as the \emph{Yukawa charge} of the solution. Once the parameters $M$ and $S_2^-$ are specified, the gravitational properties of the solutions at large distances are completely determined.

To classify different families of solutions, it is crucial to analyze the behavior of the metric near the origin or a metric singularity. Starting from the series expansion
\begin{equation} \label{eq|frobenius}
\begin{split}
h(r)&=\left(r-r_0\right)^t\!\left[\displaystyle\sum_{n=0}^N h_{t+\frac{n}{\Delta}}\left(r-r_0\right)^{\frac{n}{\Delta}}\!+O\!\left(\left(r-r_0\right)^{\frac{N\!+\!1}{\Delta}}\right)\!\right],\\
f(r)&=\left(r-r_0\right)^s\!\left[\displaystyle\sum_{n=0}^N f_{s+\frac{n}{\Delta}}\left(r-r_0\right)^{\frac{n}{\Delta}}\!+\!O\left(\left(r-r_0\right)^{\frac{N\!+\!1}{\Delta}}\right)\!\right],
\end{split}
\end{equation}
it is possible to use a generalized Frobenius method to determine the allowed leading exponents $(s,t)$, the parameter $\Delta$ which determines whether the power series contains integer or fractional powers, and whether the expansion is taken around a finite radius or the origin. The families of solutions can then be classified by the label $(s,t)_{r_0}^\Delta$~\cite{Lu:2015psa,Podolsky:2019gro,Silveravalle:2023lnl,Silveravalle:2022wij}. The relevant families of solutions are:

\paragraph{$\mathbf{(1,1)_{r_H}^1}$ - black holes.} A simultaneous root of the functions $h(r)$ and $f(r)$ defines an event horizon. In Einstein-Weyl gravity, black hole solutions have standard horizons, with the metric functions expandable in integer powers starting at the linear term. Besides the Schwarzschild solution, there are also black holes with non-Schwarzschild metrics. Both types of solutions depend on a single free parameter and, in the $(M,S_2^-)$ plane, they form two distinct lines: one for Schwarzschild black holes and one for the non-Schwarzschild ones, which intersect at a specific mass. For further details, see~\cite{Bonanno:2019rsq,Lu:2015cqa,Goldstein:2017rxn}.

\paragraph{$\mathbf{(1,0)_{r_T}^2}$ - non-symmetric wormholes.} A root of the function $f(r)$ that is not a root of $h(r)$ defines a wormhole throat. In Einstein–Weyl gravity, the metric near the throat must be expanded in half-integer powers. This indicates that the throat is non-symmetric, and it connects an asymptotically flat spacetime to one where the metric vanishes asymptotically and a singularity lies at infinity. These solutions depend on two free parameters and, in the $(M,S_2^-)$ plane, they fill most of the region with positive mass and an attractive Yukawa charge ($S_2^-<0$). For further details, see~\cite{Lu:2015psa,Bonanno:2022ibv,perkinsthesis}.

\paragraph{$\mathbf{(\text{-}1,\text{-}1)_{0}^1}$ - repulsive (standard) naked singularities.} One of the allowed behaviors near the origin for the metric of vacuum solutions is that both $h(r)$ and $f(r)$ diverge as $r^{-1}$. This is the same behavior as in the Schwarzschild metric, but in Einstein–Weyl gravity it can also occur for naked singularities with a repulsive gravitational potential at the origin. It appears that for such naked singularities, the power series expansion must be refined with logarithmic corrections, which, however, do not alter the qualitative features of the solutions~\cite{Silveravalle:2023lnl}. These solutions depend on two free parameters and, in the $(M,S_2^-)$ plane, they fill most of the region with negative mass, as well as the region with positive mass and repulsive Yukawa charge ($S_2^->0$). For further details, see~\cite{Lu:2015psa,Silveravalle:2022wij,perkinsthesis}.

\paragraph{$\mathbf{(\text{-}2,2)_{0}^1}$ - attractive (veiled) naked singularities.} The other allowed behavior near the origin for the metric of vacuum solutions is that $h(r)$ vanishes as $r^2$ and $f(r)$ diverge as $r^{-2}$, making all terms in the metric vanish as $r^2$. This behavior is specific to quadratic gravity and corresponds to naked singularities with an attractive gravitational potential. These solutions depend on two free parameters and, in the $(M,S_2^-)$ plane, they occupy small regions near the black hole lines. This family of solutions has been extensively studied by Holdom as black hole mimickers. For further details, see~\cite{Lu:2015psa,perkinsthesis,Holdom:2002xy,Holdom:2016nek}. As anticipated in the introduction, this family of solution will be the focus of this work.

\section{From $\mathbf{(\text{-}2,2)}$ solutions to veiled singularities}

The solutions we consider are defined by the absence of an event horizon or a wormhole throat, and by the metric behaving as
\begin{equation}
    \mathrm{d}s^2\sim r^2\left(-h_{2}\mathrm{d}t^2+\frac{\mathrm{d}r^2}{f_{-2}}+\mathrm{d}\Omega^2\right)
\end{equation}
near the origin, with $h_{2}$ and $f_{-2}$ being constants. These solutions have been referred to as $(2,2)$ solutions~\cite{Lu:2015psa}, $(2,2)$-holes~\cite{Holdom:2016nek}, Bachian singularities~\cite{Podolsky:2019gro}, Holdom stars~\cite{Silveravalle:2022wij}, or attractive naked singularities~\cite{Silveravalle:2023lnl,DelPorro:2025wts}; however, we believe that a complete analysis of these solutions as viable physical objects is still lacking. In particular, the extensive work of Bob Holdom~\cite{Holdom:2002xy,Holdom:2016nek,Holdom:2019bdv,Holdom:2020onl,Holdom:2022zzo,Holdom:2022fsm} has focused on a thin-shell model, which can be tuned to mimic a black hole, rather than on the pure vacuum solutions of quadratic gravity. In this section, we provide a comprehensive overview of these solutions and show how the $(-2,2)_{0}^1$ behavior at the origin globally gives rise to a \emph{veiled singularity} spacetime. In this and following sections, we made every quantity dimensionless using the mass $m_2$ as unit.

\subsection{Characterization of the singularity}

The metric close to the origin can be expanded as
\begin{equation}\label{eq|expansion}
    \begin{split}
        h(r)=&\ h_2\,r^2\left(1-\frac{f_{-1}}{f_{-2}}r+\frac{7 f_{-1}^2+f_{-2} \left(2-6 f_0\right)}{8 f_{-2}^2}r^2+O\left(r^3\right)\right),\\
        f(r)=&\ \frac{f_{-2}}{r^2}\left(1+\frac{f_{-1}}{f_{-2}}r+\frac{f_0}{f_{-2}}r^2+O\left(r^3\right)\right),
    \end{split}
\end{equation}
where all higher-order coefficients can be expressed in terms of $h_2,\, f_{-2},\, f_{-1}$ and $f_0$. While the Ricci scalar is identically zero due to equation~\eqref{trace}, the contracted Ricci, Riemann, and Weyl tensors diverge at the origin as
\begin{equation}\label{eq|curvature}
    \begin{split}
        R^{\mu\nu}R_{\mu\nu}=&\ 12 f_{-2}^2\ r^{-8}+O\left(r^{-7}\right),\\
        R^{\mu\nu\rho\sigma}R_{\mu\nu\rho\sigma}=&\ 24 f_{-2}^2\ r^{-8}+O\left(r^{-7}\right),\\
        C^{\mu\nu\rho\sigma}C_{\mu\nu\rho\sigma}=&\ \frac{3}{16} \left(2\left(1+f_0\right)-\frac{f_{-1}^2}{f_{-2}}\right)^2r^{-4}+O\left(r^{-3}\right),
    \end{split}
\end{equation}
indicating the presence of a strong curvature singularity at the origin. However, it is worth noting that although all curvature invariants diverge, the Weyl tensor remains integrable. Including the measure, the integral of the squared Weyl tensor reads
\begin{equation}\label{eq|intweyl}
\begin{split}
    \int\mathrm{d}^4x\sqrt{-g}C^{\mu\nu\rho\sigma}C_{\mu\nu\rho\sigma}=&\, \int\mathrm{d}^4x\left[\sqrt{\frac{h_2}{f_{-2}}}r^4\sin(\theta)\frac{3}{16} \left(2\left(1+f_0\right)-\frac{f_{-1}^2}{f_{-2}}\right)^2r^{-4}+O(r)\right]\\
    =&\,\int\mathrm{d}t\left[\frac{3}{4}\pi\sqrt{\frac{h_2}{f_{-2}}} \left(2\left(1+f_0\right)-\frac{f_{-1}^2}{f_{-2}}\right)^2 r+O\left(r^2\right)\right],
    \end{split}
\end{equation}
showing that, while the spacetime possesses a curvature singularity, its action remains finite. A geodesic with energy $E$ and angular momentum $L$ reaches the singularity in coordinate and proper time given by~\cite{Silveravalle:2023lnl}
\begin{equation}\label{eq|time}
\begin{split}
    \Delta t=&\ \sqrt{\frac{E^2}{h_2f_{-2}\left(E^2-h_2L^2\right)}}\ r+O\left(r^2\right),\\
        \Delta\tau=&\ \sqrt{\frac{h_2}{f_{-2}\left(E^2-h_2L^2\right)}}\ \frac{r^3}{3}+O\left(r^4\right),
\end{split}
\end{equation}
whenever expansion~\eqref{eq|expansion} holds, independently of whether the geodesic is timelike or null (in the latter case, the proper time $\tau$ must be interpreted as an affine parameter). Finally, considering nearby geodesics in a congruence rotating in the equatorial plane, the displacement vectors behave as
\begin{equation}\label{eq|congruence}
    z_t^\mu\sim\begin{pmatrix}
        c_1\log(r)\\
        -c_1\sqrt{\frac{h_2f_{-2}\left(E^2-h_2L^2\right)}{E^2}}\log(r)\\
        -c_1\frac{h_2L}{E}\log(r)\\
        c_2r
    \end{pmatrix},\qquad
    z_n^\mu\sim\begin{pmatrix}
        c_1\\
        c_2\\
        C_3\left(c_1,c_2,E,L,h_2,f_{-2}\right)L\\
        c_4r
    \end{pmatrix},
\end{equation}
in the timelike and null cases, respectively. While the congruence of null geodesics remains relatively well-behaved, timelike geodesics tend to become infinitely separated as they approach the origin; in other words, a timelike observer would be disrupted by diverging tidal forces. Taken together, Eqs.~(\ref{eq|curvature}--\ref{eq|congruence}) demonstrate that $(-2,2)_0^1$ solutions are naked singularities.

Nevertheless, while falling into the singularity has catastrophic consequences, the possibility of a particle escaping from it must be considered with care. In particular, recall that the redshift of a photon emitted at radius $r$ and measured at infinity is $z=h(r)^{-1/2}-1$. Using the expansion~\eqref{eq|expansion}, we obtain
\begin{equation}\label{eq|redshift}
    z=\frac{1}{\sqrt{h(r)}}-1\sim\frac{1}{\sqrt{h_2}\ r}\ \xrightarrow[r\to 0]{}\ \infty.
\end{equation}
A photon emitted at the singularity therefore reaches spatial infinity with zero energy, or, equivalently, only a particle with infinite energy could escape. The singularity is in fact a degenerate Killing horizon for the time-like Killing vector $\xi^\mu\partial_\mu=\partial_t$, and no information can genuinely escape it. These peculiar features are due to the fact that their gravitational field remains attractive, in contrast with the standard naked singularities of General Relativity, where the field becomes repulsive and leads to the emission of high-energy photons. While being naked singularities from a causal point of view, $(-2,2)_0^1$ solutions are protected by a \emph{weaker} form of the cosmic censorship conjecture, in which no particle with finite energy can leave the singularity, and are therefore \emph{veiled}.

\subsection{Beyond Frobenius analysis and veil radius}

\begin{figure}[htb]
\centering
\includegraphics[width=\textwidth]{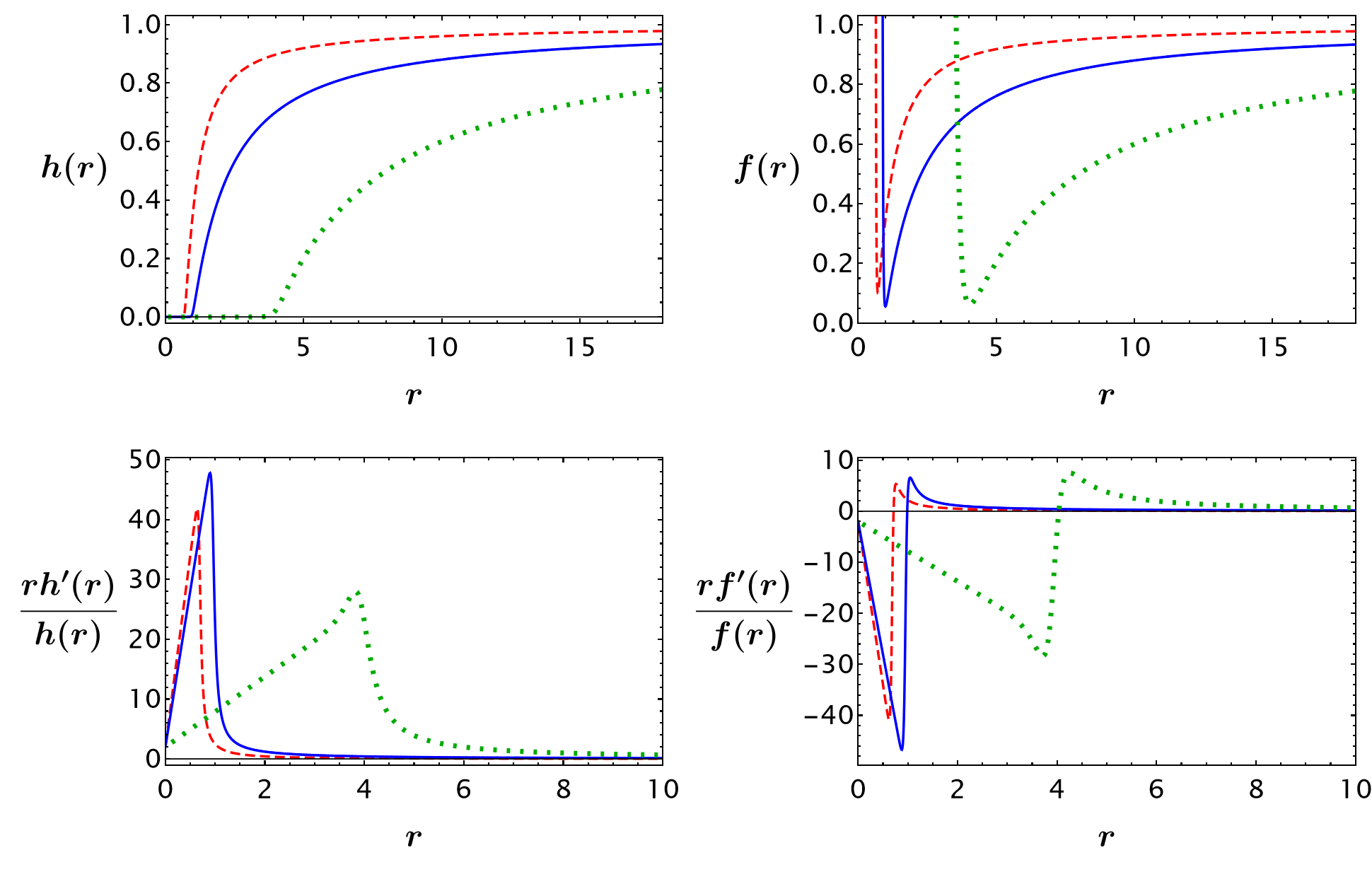}\\
\caption{In the top panels we plot the metric functions $h(r)$ and $f(r)$ for three different parameter choices, with the blue solid line corresponding to $M=0.6$ and $S^-_2=0.05$, the red dashed line to $M=0.2$ and $S^-_2=-0.2$, and the green dotted line to $M=2.0$ and $S^-_2=0.01$. The bottom panels show the logarithmic derivatives $\frac{rh'(r)}{h(r)}$ and $\frac{rf'(r)}{f(r)}$, which approach $2$ for $h(r)$ and $-2$ for $f(r)$ as $r\to 0$, in agreement with the expected near-origin scaling.}
\label{fig|metric}
\end{figure}

While the expansion~\eqref{eq|expansion} is very useful to characterize the properties of the singularity, it is not an accurate description of the full solution. In the top panels of Fig.~\ref{fig|metric} we show the two metric functions, $h(r)$ and $f(r)$, for three different veiled singularities, and it is clear that expansion~\eqref{eq|expansion} fails to reproduce their behavior in the region surrounding the singularity. In particular, the lapse function $h(r)$ drops to zero already at radii close to the Schwarzschild value $r\sim 2M$, and appears to remain vanishing all the way to the origin. The radial metric function $f(r)$, by contrast, exhibits a sharp spike near the same radius and continues to grow as $r$ decreases.

Although this might resemble a numerical artifact, the log-log derivatives of the metric functions $\frac{\mathrm{d}\log h(r)}{\mathrm{d}\log r}=\frac{rh'(r)}{h(r)}$ and $\frac{\mathrm{d}\log f(r)}{\mathrm{d}\log r}=\frac{rf'(r)}{f(r)}$ shown in the bottom panels of Fig.~\ref{fig|metric} confirm that the Frobenius expansion in~\eqref{eq|expansion} is valid only extremely close to the origin, and that exponential corrections must be included.


Considering the region where the log-log derivatives can be approximated linearly, the metric functions behave as
\begin{equation}
    \begin{array}{cc}
         & \frac{rh'(r)}{h(r)}\sim2+a\,r  \\[0.3cm]
         & \frac{rf'(r)}{f(r)}\sim-2-a\,r 
    \end{array}
    \qquad\implies\qquad
    \begin{array}{cc}
         & h(r)\sim h_2 r^2\,\mathrm{e}^{a\,r}  \\[0.3cm]
         & f(r)\sim \frac{f_{-2}}{r^2}\,\mathrm{e}^{-a\,r}
    \end{array}
\end{equation}
with $a,\, h_2$ and $f_{-2}$ as free parameters. This form still does not provide the full set of integration constants, it matches Eq.~\eqref{eq|expansion} only at next-to-leading order, and it does not satisfy the equations of motion, but it reproduces the qualitative behavior of the solution. An exponential dependence of this kind already appeared in the asymptotic region of non-symmetric wormholes. More generally, expanding around a vanishing metric as
\begin{equation}
    h(r)=\epsilon\, h_1(r)\left(1+\epsilon\, h_2(r)+O\left(\epsilon^2\right)\right),\qquad f(r)=\frac{1}{\epsilon\, f_1(r)\left(1+\epsilon\, f_2(r)+O\left(\epsilon^2\right)\right)},
\end{equation}
one finds at leading order in $\epsilon$, that is $O\left(\epsilon^{-1}\right)$, the solutions
\begin{equation}\label{eq|expo}
    h_1(r)=h_2 r^2\,\mathrm{e}^{a\,r},\qquad\qquad f_1(r)= \frac{f_{-2}}{r^2}\,\mathrm{e}^{-a\,r},
\end{equation}
while at the next order the missing parameters are recovered~\cite{Bonanno:2022ibv}. Non-symmetric wormholes appear for negative values of $a$ in the $r\to\infty$ region, while veiled singularities appear for positive values of $a$ in the $r\to 0$ region.

We can then define the veil radius as the largest radius up to which the metric can be reliably approximated by~\eqref{eq|expo}:
\begin{equation}\label{eq|effective}
    r_{\rm v}:= \max\left\{r\ |\ \left|\frac{rh'(r)}{h(r)}-(2+a\,r)\right|<\delta\ \land\ \left|\frac{rf'(r)}{f(r)}+(2+a\,r)\right|<\delta\right\},
\end{equation}
with $\delta$ small (in practice we take $\delta=10^{-3}$). Below this radius, the lapse function is essentially zero and the radial function $f(r)$ becomes very large, due to the smallness of $h_2$ (of order $10^{-15}$–$10^{-20}$) and the largeness of $f_{-2}$ (of order $10^{10}$–$10^{15}$). Within this approximation, the curvature invariants behave as
\begin{equation}\label{eq|curvaturenf}
    \begin{split}
        R^{\mu\nu}R_{\mu\nu}\sim &\ 2\,\mathrm{e}^{-2\,a\,r}\left(6+4\,a\,r+a^2r^2\right) f_{-2}^2\ r^{-8},\\
        R^{\mu\nu\rho\sigma}R_{\mu\nu\rho\sigma}\sim &\ 4\,\mathrm{e}^{-2\,a\,r}\left(6+4\,a\,r+a^2r^2\right) f_{-2}^2\ r^{-8},\\
        C^{\mu\nu\rho\sigma}C_{\mu\nu\rho\sigma}\sim &\ 0,
    \end{split}
\end{equation}
so both the Ricci and Riemann contractions are large, and the spacetime is effectively in a regime of extreme curvature. 
Geodesics falling to the origin reach it in coordinate and proper time
\begin{equation}\label{eq|timenf}
\begin{split}
        \Delta t\sim &\ \frac{1}{\sqrt{h_2f_{-2}}}\ r,\\
        \Delta\tau\sim &\ \sqrt{\frac{h_2}{f_{-2}}}\ \frac{\mathrm{e}^{a\,r}\left(2-2a\,r+a^2r^2\right)}{a^3E},
\end{split}
\end{equation}
where the angular momentum has no significant role, indicating that trajectories are essentially radial. The coordinate time is unaffected by the exponential factor, and the combination $h_2f_{-2}$ ensures it remains of moderate size; the proper time, instead, quickly tends to zero because $h_2/f_{-2}$ is very small. This mismatch already signals a large redshift for photons emitted inside $r_{\rm v}$ and received at infinity:
\begin{equation}\label{eq|redshiftnf}
    z=\frac{1}{\sqrt{h(r)}}-1\sim\frac{1}{\sqrt{h_2}\ r\ \mathrm{e}^{a\,r}}-1 \gg 1,\qquad\text{if }\ h_2 \ll 1.
\end{equation}
Radiation from inside the veil thus reaches infinity with negligible energy and is effectively hidden. Finally, although the geometry is extreme, singular behavior occurs only at the origin. The displacement vectors of geodesic congruences,
\begin{equation}\label{eq|congruence2}
    z_t^\mu\sim\begin{pmatrix}
        c_1+c_2\log(r)+\frac{aEc_2-2Lc_3}{2E}r-\frac{aLc_3}{4E}r^2\\
        \sqrt{h_2f_{-2}}\left((c_2-c_1)-c_2\log(r)-\frac{ac_2}{2}r+\frac{aLc_3}{4E}r^2\right)\\
        c_3r\\
        c_4r
    \end{pmatrix},\qquad
    z_n^\mu\sim\begin{pmatrix}
        c_1-\frac{Lc_3}{2E}r\\
        c_2-\frac{\sqrt{h_2f_{-2}}Lc_3}{2E}r\\
        c_3r\\
        c_4r
    \end{pmatrix},
\end{equation}
shows a divergent behavior only in the limit $r\to 0$.

To conclude, we note that in the limit of large masses veiled singularities have a small Yukawa charge $S_2^-$~\cite{Silveravalle:2022wij}, and the veil radius lies very close to the Schwarzschild radius, making these spacetimes natural black hole mimickers.

\subsection{Photon orbits and shadow}

A visualization of veiled singularities can be obtained with a ray-tracing simulation, in which light rays are traced along null geodesics. An important property to determine in such simulations is whether the spacetime admits a photon ring, i.e. a radius at which light can orbit on closed circular null geodesics. In general, spacetimes either have a single photon ring together with an event horizon—--so that light rays entering the photon ring inevitably fall into the horizon, producing a shadow—--or they have no horizons, and then two photon rings appear: one stable and one unstable, leading to concentric light-ring features. Veiled singularities behave differently in this respect.

The null geodesic equation and the null line element for a circular (i.e. with $\mathrm{d}r=0$) orbit in the equatorial plane (i.e. with $\mathrm{d}\theta=0$) are
\begin{equation}\label{eq|orbit}
        \frac{1}{2}h'(r)\left(\frac{\mathrm{d}t}{\mathrm{d}\lambda}\right)^2-r\left(\frac{\mathrm{d}\phi}{\mathrm{d}\lambda}\right)^2= 0,\qquad\qquad-h(r)\mathrm{d}t^2+r^2\mathrm{d}\phi^2= 0,
\end{equation}
which imply that a photon ring occurs at radii satisfying \cite{Chandrasekhar:1985kt,Cardoso:2008bp,Konoplya:2019sns}
\begin{equation}\label{eq|lightring}
    \left.rh'(r)-2h(r)\right|_{r=r_{\rm pr}}=0,\qquad\implies\qquad\left.\frac{rh'(r)}{h(r)}\right|_{r=r_{\rm pr}}=2.
\end{equation}
As shown in the lower-left panel of Fig.~\ref{fig|metric}, equation~\eqref{eq|lightring} is satisfied in veiled singularity spacetimes at a radius outside the veil and again at the origin; the second light ring is thus degenerate with the singularity. For spacetimes with two photon rings, light rays that cross the outer one are driven inward, but after crossing the second they are turned back by their angular momentum and return to infinity. In veiled singularities, this does not happen: light rays that cross the outer photon ring are inevitably drawn into the singularity and terminate there. Consequently, light entering the outer photon ring cannot escape, and a shadow is expected.

Assuming the weak-field expansion~\eqref{eq|weakfield}, the photon-ring radius is given by
\begin{equation}\label{eq|lightringwf}
    \left.r-3M+S_2^-\left(3+m_2\,r\right)\mathrm{e}^{-m_2\,r}\right|_{r=r_{\rm pr}}=0,
\end{equation}
which reduces to the standard $r_{\rm pr}=3M$ when $S_2^-=0$. One finds $r_{\rm pr}<3M$ for $S_2^->0$, and $r_{\rm pr}>3M$ for $S_2^-<0$; for veiled singularities of large mass, which carry a positive Yukawa charge~\cite{Silveravalle:2022wij}, the shadow will be slightly smaller than that of a Schwarzschild black hole of the same mass.

\begin{figure}[htb]
    \centering
\includegraphics[width=\textwidth]{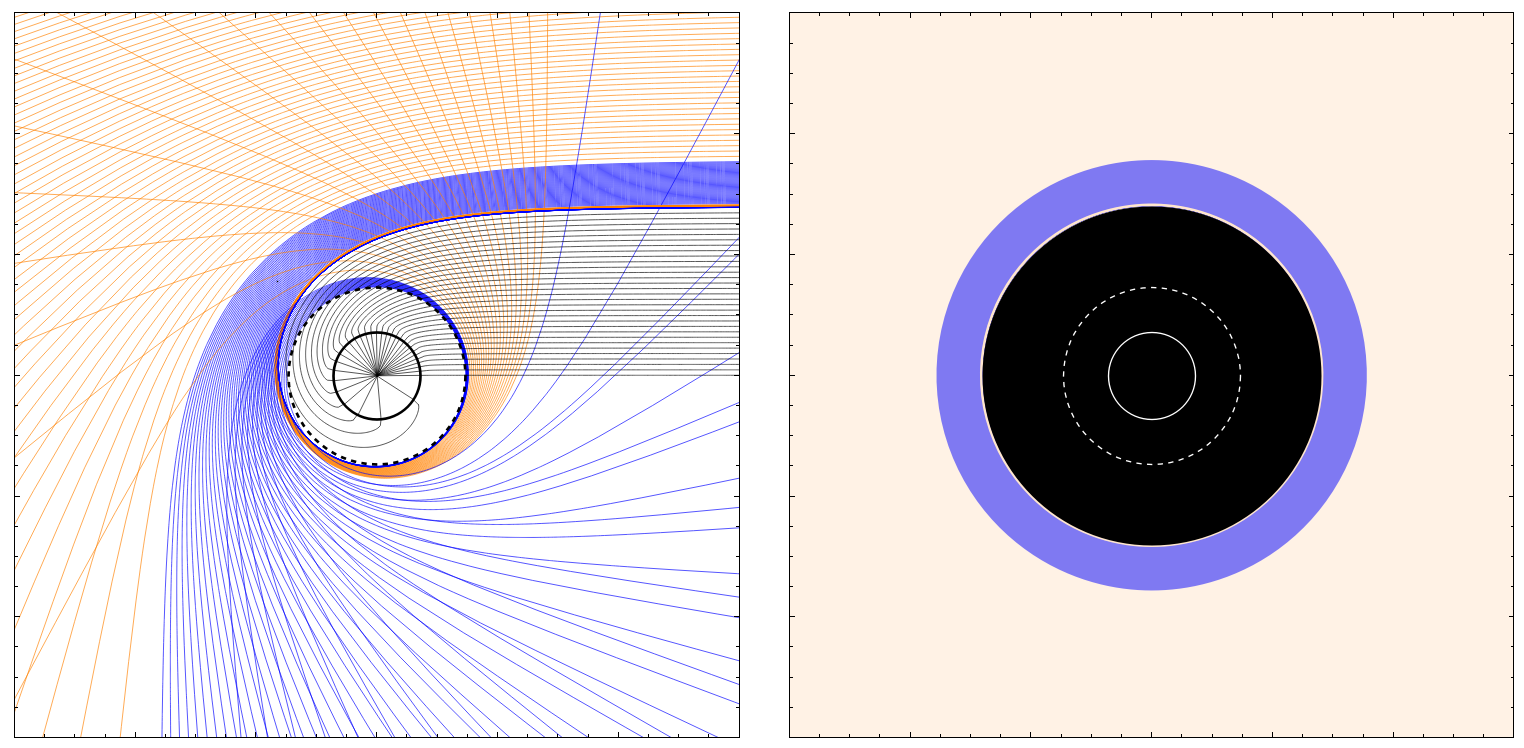}
    \caption{Null geodesics around a naked singularity in arbitrary units (for reference, we used $M=0.6$ and $S_2^-=0.07$). Left panel illustrates a ray-tracing simulation, with an observer at $x=-\infty$ and an illuminated screen assumed at $x=+\infty$. Orange lines are light rays that reach the observer, blue lines are light rays deflected back by strong lensing effects. and black lines are light rays that reach the singularity. The dashed and solid circles indicate the position of the external photon ring and the veil radius. Right panel illustrate the face-on view of the singularity from the observer position.}
    \label{fig|raytracing}
\end{figure}

Finally, in the left panel of Fig.~\ref{fig|raytracing} we show the ray-tracing simulation, where the observer is placed at $x=-\infty$ and an illuminated screen at $x=+\infty$, and in the right panel we show the face-on view of the veiled singularity as seen by the observer. Light rays that pass sufficiently far from the photon ring are bent but still reach the observer. As the impact parameter decreases, the deflection grows, and beyond a certain point the rays are bent back to the illuminated screen, never reaching the observer. Rays that pass very close to the photon ring alternate between being deflected toward the observer and back to the screen, producing the observable light ring. Those that cross inside the photon ring are inevitably captured by the singularity; approaching the veil, they become effectively radial and end at the singularity. From the face-on view, the veiled singularity appears with a shadow of radius $r_{\rm sh}=r_{\rm pr}h(r_{\rm pr})^{-1/2}$. Its image is thus the same as that of a Schwarzschild black hole, except that here $r_{\rm sh}\neq 3\sqrt{3}M$.

To conclude, veiled singularities would appear qualitatively as black holes, though with quantitative differences that make them, at least in principle, observationally distinguishable.

\section{Stability}

The purpose of this section is to assess the stability of the solutions discussed above under linear perturbations. In particular, we focus on massive tensor perturbations, which are peculiar to this theory due to the presence of a massive spin-2 ghost mode.

In general studying linear perturbation corresponds to consider the metric as
\begin{equation}
    g_{\mu\nu}=g_{\mu\nu}^0+{\epsilon}h_{\mu\nu},
\end{equation}
that is, as the sum of the background metric $g_{\mu\nu}^0$ plus a small perturbation. This leads to an expansion of the equations of motion~\cite{Silveravalle:2023lnl}. One of the most important results obtained in this field, due to Regge, Wheeler and Zerilli~\cite{Regge:1957td,Zerilli:1970se}, is the reduction of these tensor equations, for spherically symmetric backgrounds, to scalar wave equations of the form
\begin{equation}
    \left(\frac{\mathrm{d}^2}{\mathrm{d}t^2}-\frac{\mathrm{d}^2}{\mathrm{d}r^2_*}\right)\phi(r,t)+V(r)\phi(r,t)=0  \label{pert},
\end{equation}
where $r_*$ is the tortoise coordinate, defined 
as
\begin{equation}
    r_*=\int{\frac{\mathrm{d}r}{\sqrt{h(r)f(r)}}}.
\end{equation}
The general form of Eq.~\eqref{pert} is independent from the specific type of perturbation; the differences arise solely from the potential $V(r)$, determined by the theory taken into consideration and the background metric. Determining the correct expression of the potential in quadratic gravity is precisely the main challenge in studying such perturbations. Fortunately, this task has been elegantly accomplished by A. Held and J. Zang~\cite{Held:2022abx} for monopole perturbation ($l=0$). Starting from the action in the Einstein frame, they derived a Zerilli-type equation with a potential of the form:
 \begin{multline} \label{pot}
    V(r)= h + \frac{f'h + fh'}{2r} + \\
\frac{288 f^3 h^3 (-2h + h'r)^2}{\left(4(1 - 3f)h^2r + (f'h + fh')(3f h'r + h(-4 + 3f'r))\right)^2} + \\
\frac{24 f h^2 (2f + f'r)(2h - h'r)}{r \left(4(1 - 3f)h^2r + (f'h + fh')(3f h'r + h(-4 + 3f'r))\right)}
\end{multline}
where all quantities are expressed in units of $m_2$.

\subsection{Numerical methods} \label{nummet}

We now want to extract information of the time evolution of the perturbation from equation~\eqref{pert}. To obtain the solution numerically, we followed the technique explained in~\cite{Konoplya:2011qq}. Introducing the light-cone coordinates $v=t+r_*$ and $u=t-r_*$, the equation~\eqref{pert} takes the form
\begin{equation}
    4\frac{\mathrm{d}^2}{\mathrm{d}v \mathrm{d}u}\phi(u,v)+V(u,v)\phi(u,v)=0.
\end{equation}
To integrate the wave equation numerically, we discretize the \((u,v)\) plane on a uniform grid with step size \(h\). The field \(\phi(u,v)\) is then evolved using the standard scheme~\cite{Konoplya:2011qq,Silveravalle:2023lnl,Spina:2024npx}
\begin{equation}\label{pertgrid}
    \phi(N)=\phi(W)+\phi(E)-\phi(S)+\frac{h^2}{8}V(S)(\phi(W)+\phi(E)),
\end{equation}
where the grid points are defined as
$$
S = (u, v), \qquad 
W = (u + h, v), \qquad 
E = (u, v + h), \qquad 
N = (u + h, v + h).
$$
With appropriate initial data, the integration can be performed iteratively, row by row, to obtain the time-domain evolution of the perturbation $\phi$.

Practically, we developed a Fortran code to implement this procedure. Our analysis was divided into two spatial regions:  
(i) large distances in the asymptotic weak-field regime, starting from a point \(x_1\), where we used Eq.~\eqref{eq|weakfield}; and  
(ii) from \(x_1\) down to a point close to the origin, where the equations were solved numerically using Eq.~\eqref{eom}, with the weak-field solution~\eqref{eq|weakfield} providing the boundary conditions.  
This integration was carried out using the integrator packet \texttt{DO2PDF} implemented by NAG group (see \href{https://www.nag.com}{https://www.nag.com})~\cite{Press:1992zz}. 

\begin{figure}[htb]
\centering
\includegraphics[width=0.6\textwidth]{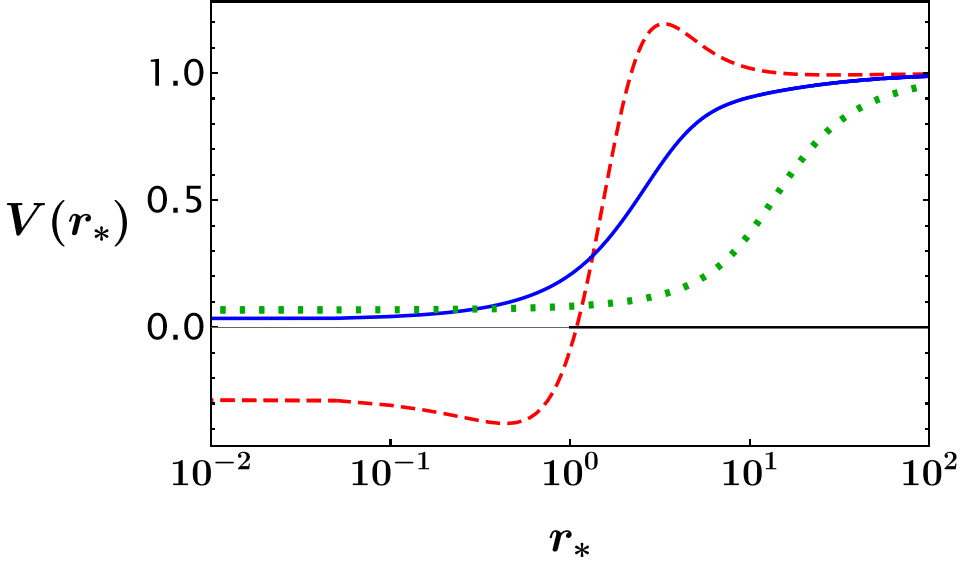}
\caption{Effective potential as a function of the tortoise coordinate for different values of $M$ and $S_2^-$. Blue solid line: $M=0.6$ and $S^-_2=0.05$; red dashed line: $M=0.2$ and $S^-_2=-0.2$; green dotted line: $M=2.0$ and $S^-_2=0.01$. In all cases, we have a potential well with a finite value in the origin.}
\label{potential}
\end{figure}

Particular care must be taken with the potential, since it must be expressed as a function of the tortoise coordinate, whereas in~\eqref{pot} it is as a function of $r$. The inversion of the tortoise coordinate was implemented by integrating, together with the equations of motion, the additional differential relation
\begin{equation}\label{eq:tortoisediff}
    \frac{\mathrm{d}r_*(r)}{\mathrm{d}r} = \frac{1}{\sqrt{h(r)f(r)}}.
\end{equation}
For each value of the tortoise coordinate, the corresponding radius \(\bar{r}(r_*)\) is then obtained by finding the zero of
\begin{equation}\label{eq:tortoiseinv}
    r_*(r) - r_* = 0,
\end{equation}
where $r_*(r)$ denotes the numerical integration of the tortoise coordinate. Having a potential that depends solely on the radius simplifies the procedure, allowing us to compute the potential over the full range of values of \(r_*\) at the start of the integration. 
After performing this inversion, we obtain the potential expressed in terms of the metric functions in the two spatial regions (see Fig.~\ref{potential}).

For the initial conditions, there is no unique prescription, but a common choice is to impose a Gaussian perturbation along the $u=0$ axis,
\begin{equation} \label{gauss}
    \phi(v,0)=e^{-\frac{(v-v_0)^2}{2\sigma}}
\end{equation}
which represents an initial impulse of the perturbation with a generic form that does not affect the later-time behavior. The second condition is imposed at the singularity ($r_*=0, u=v)$ where we set the perturbation to zero, since we consider the point lying outside the spacetime, thus, a perturbation there has no physical meaning. 
After imposing these conditions, we integrated Eq.~\eqref{pertgrid} over the grid and extracted the time evolution of the perturbation at a fixed radius $x_{\rm meas}$. To ensure the robustness of our analysis, we performed several consistency tests, varying the grid step size approaching the singularity at the origin, and evaluating the time evolution at different fixed radii. In each of these cases, the overall qualitative behavior remained unchanged.

\subsection{Results} \label{results} 

\begin{figure}[h] 
\centering
\includegraphics[width=\textwidth]{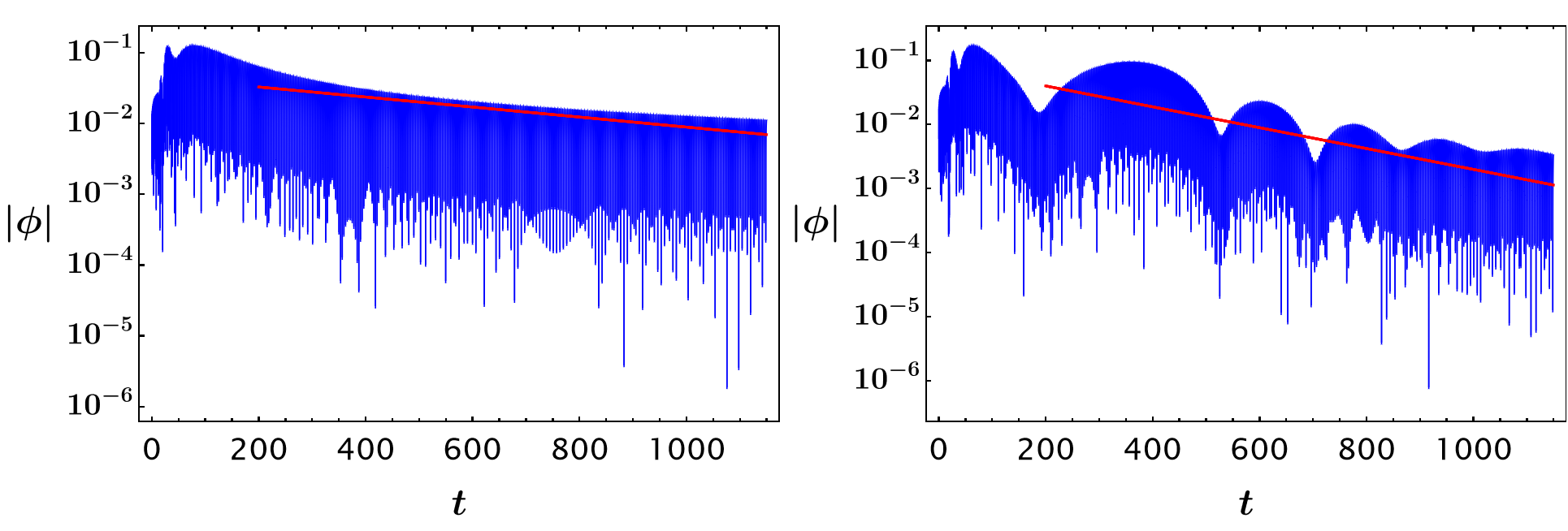}
\caption{Two different examples of the time evolution of the perturbations with the corresponding linear fit, on the left with $M=0.6$ and $S_2^-=0.05$, on the right with $M=0.2$ and $S_2^-=-0.2$ showing characteristic beating patterns.}
\label{timedom}
\end{figure}

The time evolution of the perturbation provides a direct tool to explore the stability of the phase diagram. To achieve this goal, we employed two complementary methods to assess stability. First, we performed a simple linear fit of the logarithm of the perturbation amplitude using the standard Linear Least Squares method. The fit was carried out after the first peak of the signal, since the position in the time of this peak depends on the initial conditions, in particular on the location of the Gaussian pulse used as initial perturbation, and on the observation radius at which the perturbation is evaluated.
By analyzing the general trend, we assess stability through the slope coefficient of the fitted line, which provides an approximation of the decay rate, i.e., the imaginary part \(\omega_{\mathrm{Im}}\) of the perturbation frequency (see Fig.~\ref{timedom}).

As a consistency check, we independently verified the behavior using Prony’s method, which is specifically designed to extract the frequencies of signals composed of damped complex exponentials. In this approach, the data are fitted by a linear combination of terms of the form
\begin{equation} \label{prony}
    \phi(t)\simeq \sum_{i=1}^p C_i e^{-i\omega_i t}
\end{equation}   
from which the frequencies can be obtained as~\cite{Konoplya:2011qq,pozrikidis2011introduction,Berti:2007dg},
\begin{equation}
    \omega_j=\frac{i}{h}\log z_j
\end{equation}
where $h$ is the time sampling interval, and $z_j$ are the roots of the polynomials derived from Eq.~\eqref{prony}.  

\begin{figure}[h] 
\centering
\includegraphics[width=0.8\textwidth]{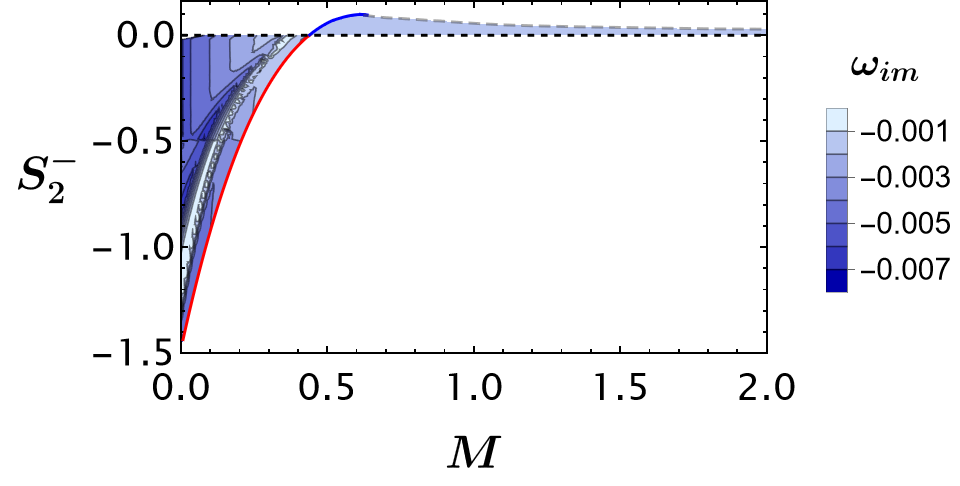}
\caption{Diagram of the stability as a function of the parameters. The colors scale represents the variation of the imaginary part of the perturbations frequencies. Solid red and blue lines correspond to non-Schwarzschild black holes, while the dashed black line represents Schwarzschild black holes.}
\label{griglia}
\end{figure}

Using these two methods, we explored the stability of the solutions with $M>0$ in the phase diagram. In particular, we constructed a grid with step size of $0.01$ in the parameter space and extracted approximate values of the imaginary part of the perturbation frequencies via the linear fit. These results were then used to generate the contour plot shown in Fig.~\ref{griglia}.
The most important outcome is that all modes are stable, i.e., they exhibit a negative imaginary part of the frequency $\omega_{\rm Im}$. This implies that these configurations can exist as physically viable solutions and, in principle, could exist in nature.

Moreover, for $S_2^->0$ the behavior remains approximately constant as the mass increases, corresponding to the most relevant regime in an astrophysical context. In contrast, when $S_2^- < 0$, we observe a lighter band in the stability map near the border with the non-Schwarzschild black holes. In this region, the signal resembles the one shown in the right panel of Fig.~\ref{timedom}, but in these cases the beat pattern is more stretched, resulting in only a single oscillation within the integration time interval, consequently the peak of the beat conditions the fit and his coefficients, yielding slightly higher values of the frequencies, though still negative, since the beats decay over time. These characteristic echoes resemble beating patterns familiar from other areas of physics, such as acoustic, so we interpret them as a consequence of the boundary conditions imposed at the singularity. The vanishing of the perturbation there introduces an effective node, which under certain conditions leads to a partial overlap of modes and the appearance of beats.  

\begin{figure}[htb] 
\centering
\includegraphics[width=\textwidth]{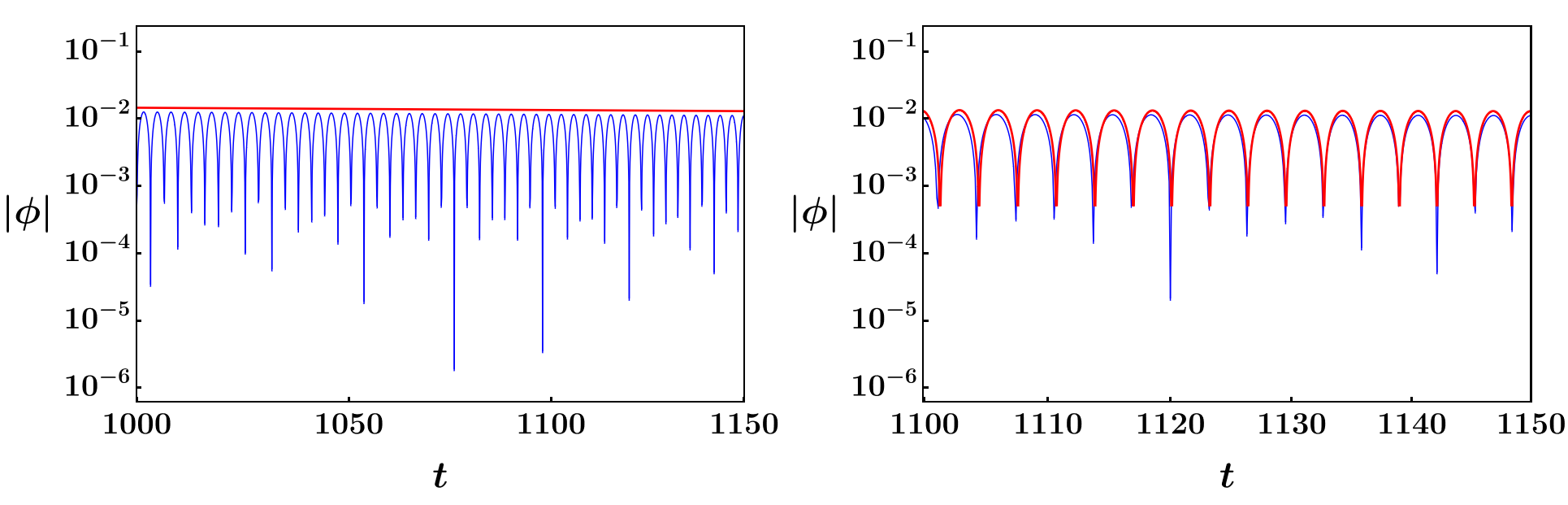}
\caption{Left panel: logarithmic plot of the asymptotic tail for $M=0.6$ and $S_2^-=0.05$. The red line corresponds to a power-law fit $\propto t^{-5/6}$.\\
Right panel: logarithmic plot of the oscillation amplitude of the asymptotic tail, fitted with the function $\propto \sin(m_2 t - \delta)$.}
\label{Tail}
\end{figure}

Having established the time evolution of the perturbations and confirmed their stability, we proceed to analyze the late-time behavior of the signal, namely the asymptotic tail, and compare it with the known behavior of black holes in classical General Relativity (GR)~\cite{Price:1971fb,Price:1972pw} and in quadratic gravity~\cite{xhtc-9cf4}. The results are shown in Fig.~\ref{Tail}, where the left panel displays the power-law decay of the tail, while the right panel shows the fit of its oscillatory part.

Unlike classical massless field perturbations, where the asymptotic tails are non-oscillatory and typically emerge only after the signal has decayed by six or more orders of magnitude, the tails observed here are oscillatory and decay more slowly. Specifically, the late-time behavior of the perturbation is well described by the following expression:
\begin{equation}
    \phi(t) \propto \sin(m_2 t - \delta)\, t^{-5/6}.
\end{equation}
This behavior is consistent with previous analyses of massive perturbations~\cite{Konoplya:2006gq,Konoplya:2023fmh,Koyama:2000hj}, and in particular with studies of the same type of perturbations in black holes within quadratic gravity~\cite{xhtc-9cf4}. These results provide further evidence of the universal power-law decay of asymptotic tails for massive perturbations and highlight the similarities between these veiled singularities and their black hole counterparts in quadratic gravity.

\section{Conclusions}

In this work we studied the physical properties of a specific class of solutions of quadratic gravity, previously known in the literature as $(2,2)$ solutions, $(2,2)$-holes, Bachian singularities, Holdom stars, or attractive naked singularities, which we propose to refer to as \emph{veiled singularities}. These configurations are characterized by the presence of a naked curvature singularity at the origin where, however, the gravitational potential remains attractive---in contrast with standard naked singularities (such as the Reissner–Nordstr\"om or Kerr ones) where it becomes repulsive.

We analyzed the geometrical structure near the origin and showed that, while the curvature diverges and the dynamics of geodesic congruences is singular, the presence of a degenerate Killing horizon at the singularity prevents any information from escaping, thus concealing it through a \emph{weaker} form of cosmic censorship. We also showed that the specific form of the metric near the origin ensures that, although the curvature is divergent, the integral of the squared Weyl tensor remains finite, making the total action finite as well.

We extended the analysis to the global geometry and found that the metric exhibits extreme properties inside a characteristic radius $r_{\rm v}$, which we called the veil radius, whose size is comparable to the would-be Schwarzschild horizon. Photons emitted within the veil experience a large redshift before reaching spatial infinity, and are thus received with negligible---possibly undetectable---energy, effectively hiding the interior region.
The attractive nature of the metric is reflected in the presence of a light ring that focuses all light rays crossing it toward the singularity. As a consequence, even though the spacetime lacks an event horizon, it still produces a shadow of size comparable to that of a Schwarzschild black hole with the same mass.

Finally, we studied the time evolution of linear perturbations of veiled singularities. In particular, we explored the region of parameter space where such configurations exist and examined whether perturbations decay with time, finding that this is always the case. In other words, veiled singularities are linearly stable, in contrast with standard naked singularities. A detailed time-domain analysis further shows that perturbations can display a distinctive beat pattern before settling into the asymptotic behavior typical of massive fields at late times.

Taken together, these results identify veiled singularities as a genuinely distinct and physically consistent class of naked singularities. Although a curvature singularity remains, they avoid the physical and observational pathologies of conventional naked singularities and are therefore no less acceptable than ordinary singular black holes as models of astrophysical compact objects. 
While the resolution of the singularity problem may require further extensions of the theory—--such as the inclusion of even higher-order derivative terms~\cite{Giacchini:2024exc,Giacchini:2025gzw}—--the quadratic corrections already open new possibilities for long-standing issues in General Relativity, for instance by providing plausible stable remnants of black hole evaporation.
Understanding whether veiled singularities can form dynamically through realistic collapse scenarios remains an open and compelling question for future investigation.

\section*{Acknowledgments}

This work was partially funded by the the INFN projects FLAG and QUAGRAP. The research of S. S. is funded by the European Union - NextGenerationEU, in the framework of the PRIN Project “Charting unexplored avenues in Dark Matter" (20224JR28W).
 
\bibliography{whbibesterna}

@article{Giacchini:2025gzw,
    author = "Giacchini, Breno L. and Kol{\'a}{\v{r}}, Ivan",
    title = "{Black holes and other exact solutions in six-derivative gravity}",
    eprint = "2503.17318",
    archivePrefix = "arXiv",
    primaryClass = "gr-qc",
    doi = "10.1103/1mkn-kfpx",
    journal = "Phys. Rev. D",
    volume = "112",
    number = "2",
    pages = "024051",
    year = "2025"
}

@article{Giacchini:2024exc,
    author = "Giacchini, Breno L. and Kol{\'a}{\v{r}}, Ivan",
    title = "{Toward regular black holes in sixth-derivative gravity}",
    eprint = "2406.00997",
    archivePrefix = "arXiv",
    primaryClass = "gr-qc",
    doi = "10.1103/PhysRevD.110.104056",
    journal = "Phys. Rev. D",
    volume = "110",
    number = "10",
    pages = "104056",
    year = "2024"
}

@article{Christodoulou:1984djm,
    author = "Christodoulou, Demetrios",
    editor = "Bertotti, B. and de Felice, F. and Pascolini, A.",
    title = "{Gravitational Collapse of a Dust Cloud and the Cosmic Censorship Conjecture}",
    doi = "10.1007/978-94-009-6469-3_3",
    journal = "Fundam. Theor. Phys.",
    volume = "9",
    pages = "27--36",
    year = "1984"
}

@article{Daas:2022iid,
    author = "Daas, Jesse and Kuijpers, Kolja and Saueressig, Frank and Wondrak, Michael F. and Falcke, Heino",
    title = "{Probing quadratic gravity with the Event Horizon Telescope}",
    eprint = "2204.08480",
    archivePrefix = "arXiv",
    primaryClass = "gr-qc",
    doi = "10.1051/0004-6361/202244080",
    journal = "Astron. Astrophys.",
    volume = "673",
    pages = "A53",
    year = "2023"
}

@article{Christodoulou:1991yfa,
    author = "Christodoulou, Demetrios",
    title = "{The formation of black holes and singularities in spherically symmetric gravitational collapse}",
    doi = "10.1002/cpa.3160440305",
    journal = "Commun. Pure Appl. Math.",
    volume = "44",
    number = "3",
    pages = "339--373",
    year = "1991"
}

@article{Penrose:1964wq,
    author = "Penrose, Roger",
    title = "{Gravitational collapse and space-time singularities}",
    doi = "10.1103/PhysRevLett.14.57",
    journal = "Phys. Rev. Lett.",
    volume = "14",
    pages = "57--59",
    year = "1965"
}

@article{Hawking:1976ra,
    author = "Hawking, S. W.",
    title = "{Breakdown of Predictability in Gravitational Collapse}",
    doi = "10.1103/PhysRevD.14.2460",
    journal = "Phys. Rev. D",
    volume = "14",
    pages = "2460--2473",
    year = "1976"
}

@article{Christodoulou:1994hg,
    author = "Christodoulou, Demetrios",
    title = "{Examples of naked singularity formation in the gravitational collapse of a scalar field}",
    doi = "10.2307/2118619",
    journal = "Annals Math.",
    volume = "140",
    pages = "607--653",
    year = "1994"
}

@article{Arrechea:2023oax,
    author = "Arrechea, Julio and Barcel{\'o}, Carlos and Carballo-Rubio, Ra{\'u}l and Garay, Luis J.",
    title = "{Ultracompact horizonless objects in order-reduced semiclassical gravity}",
    eprint = "2310.12668",
    archivePrefix = "arXiv",
    primaryClass = "gr-qc",
    doi = "10.1103/PhysRevD.109.104056",
    journal = "Phys. Rev. D",
    volume = "109",
    number = "10",
    pages = "104056",
    year = "2024"
}

@phdthesis{Arrechea:2023wgy,
    author = "Arrechea, Julio",
    title = "{Hydrostatic equilibrium in the semiclassical approximation}",
    school = "Universidad de Granada",
    month = "5",
    year = "2023"
}

@article{Arrechea:2021xkp,
    author = "Arrechea, Julio and Barcel{\'o}, Carlos and Carballo-Rubio, Ra{\'u}l and Garay, Luis J.",
    title = "{Semiclassical relativistic stars}",
    eprint = "2110.15808",
    archivePrefix = "arXiv",
    primaryClass = "gr-qc",
    doi = "10.1038/s41598-022-19836-8",
    journal = "Sci. Rep.",
    volume = "12",
    number = "1",
    pages = "15958",
    year = "2022"
}

@book{Hawking:1973uf,
    author = "Hawking, S. W. and Ellis, G. F. R.",
    title = "{The Large Scale Structure of Space-Time}",
    doi = "10.1017/CBO9780511524646",
    isbn = "978-0-521-20016-5, 978-0-521-09906-6, 978-0-511-82630-6, 978-0-521-09906-6",
    publisher = "Cambridge University Press",
    series = "Cambridge Monographs on Mathematical Physics",
    month = "2",
    year = "2011"
}

@article{Holdom:2002xy,
    author = "Holdom, Bob",
    title = "{On the fate of singularities and horizons in higher derivative gravity}",
    eprint = "hep-th/0206219",
    archivePrefix = "arXiv",
    reportNumber = "UTPT-02-09",
    doi = "10.1103/PhysRevD.66.084010",
    journal = "Phys. Rev. D",
    volume = "66",
    pages = "084010",
    year = "2002"
}

@article{Zwiebach:1985uq,
    author = "Zwiebach, Barton",
    title = "{Curvature Squared Terms and String Theories}",
    reportNumber = "UCB-PTH-85/10",
    doi = "10.1016/0370-2693(85)91616-8",
    journal = "Phys. Lett. B",
    volume = "156",
    pages = "315--317",
    year = "1985"
}

@article{Benedetti:2013jk,
    author = "Benedetti, Dario",
    title = "{On the number of relevant operators in asymptotically safe gravity}",
    eprint = "1301.4422",
    archivePrefix = "arXiv",
    primaryClass = "hep-th",
    reportNumber = "AEI-2013-032",
    doi = "10.1209/0295-5075/102/20007",
    journal = "EPL",
    volume = "102",
    number = "2",
    pages = "20007",
    year = "2013"
}

@article{Stelle:1976gc,
    author = "Stelle, K.S.",
    title = "{Renormalization of Higher Derivative Quantum Gravity}",
    reportNumber = "PRINT-76-1059 (BRANDEIS)",
    doi = "10.1103/PhysRevD.16.953",
    journal = "Phys. Rev. D",
    volume = "16",
    pages = "953--969",
    year = "1977"
}

@article{Lu:2015cqa,
    author = "Lu, H. and Perkins, A. and Pope, C.N. and Stelle, K.S.",
    title = "{Black Holes in Higher-Derivative Gravity}",
    eprint = "1502.01028",
    archivePrefix = "arXiv",
    primaryClass = "hep-th",
    reportNumber = "IMPERIAL-TP-15-KSS-01, MI-TH-1504, CAQS-1501",
    doi = "10.1103/PhysRevLett.114.171601",
    journal = "Phys. Rev. Lett.",
    volume = "114",
    number = "17",
    pages = "171601",
    year = "2015"
}

@article{Lu:2015psa,
    author = "Lü, H. and Perkins, A. and Pope, C.N. and Stelle, K.S.",
    title = "{Spherically Symmetric Solutions in Higher-Derivative Gravity}",
    eprint = "1508.00010",
    archivePrefix = "arXiv",
    primaryClass = "hep-th",
    reportNumber = "IMPERIAL-TP-15-KSS-02, MI-TH-1528",
    doi = "10.1103/PhysRevD.92.124019",
    journal = "Phys. Rev. D",
    volume = "92",
    number = "12",
    pages = "124019",
    year = "2015"
}

@article{Goldstein:2017rxn,
    author = "Goldstein, Kevin and Mashiyane, James Junior",
    title = "{Ineffective Higher Derivative Black Hole Hair}",
    eprint = "1703.02803",
    archivePrefix = "arXiv",
    primaryClass = "hep-th",
    doi = "10.1103/PhysRevD.97.024015",
    journal = "Phys. Rev. D",
    volume = "97",
    number = "2",
    pages = "024015",
    year = "2018"
}

@article{Podolsky:2018pfe,
    author = "Podolsky, Jiri and Svarc, Robert and Pravda, Vojtech and Pravdova, Alena",
    title = "{Explicit black hole solutions in higher-derivative gravity}",
    eprint = "1806.08209",
    archivePrefix = "arXiv",
    primaryClass = "gr-qc",
    doi = "10.1103/PhysRevD.98.021502",
    journal = "Phys. Rev. D",
    volume = "98",
    number = "2",
    pages = "021502",
    year = "2018"
}

@article{Podolsky:2019gro,
    author = "Podolský, Jiri and \v Svarc, Robert and Pravda, Vojtech and Pravdova, Alena",
    title = "{Black holes and other exact spherical solutions in Quadratic Gravity}",
    eprint = "1907.00046",
    archivePrefix = "arXiv",
    primaryClass = "gr-qc",
    doi = "10.1103/PhysRevD.101.024027",
    journal = "Phys. Rev. D",
    volume = "101",
    number = "2",
    pages = "024027",
    year = "2020"
}

@article{Stelle:1977ry,
    author = "Stelle, K. S.",
    title = "{Classical Gravity with Higher Derivatives}",
    reportNumber = "Print-77-0417 (BRANDEIS)",
    doi = "10.1007/BF00760427",
    journal = "Gen. Rel. Grav.",
    volume = "9",
    pages = "353--371",
    year = "1978"
}

@article{Holdom:2016nek,
    author = "Holdom, Bob and Ren, Jing",
    title = "{Not quite a black hole}",
    eprint = "1612.04889",
    archivePrefix = "arXiv",
    primaryClass = "gr-qc",
    doi = "10.1103/PhysRevD.95.084034",
    journal = "Phys. Rev. D",
    volume = "95",
    number = "8",
    pages = "084034",
    year = "2017"
}

@article{Holdom:2019bdv,
    author = "Holdom, Bob",
    title = "{Not quite black holes at LIGO}",
    eprint = "1909.11801",
    archivePrefix = "arXiv",
    primaryClass = "gr-qc",
    doi = "10.1103/PhysRevD.101.064063",
    journal = "Phys. Rev. D",
    volume = "101",
    number = "6",
    pages = "064063",
    year = "2020"
}

@article{Holdom:2020onl,
    author = "Holdom, Bob",
    title = "{Damping of gravitational waves in 2-2-holes}",
    eprint = "2004.11285",
    archivePrefix = "arXiv",
    primaryClass = "gr-qc",
    doi = "10.1016/j.physletb.2020.136023",
    journal = "Phys. Lett. B",
    volume = "813",
    pages = "136023",
    year = "2021"
}

@article{Holdom:2022zzo,
    author = "Holdom, Bob",
    title = "{2-2-holes simplified}",
    eprint = "2202.08442",
    archivePrefix = "arXiv",
    primaryClass = "gr-qc",
    doi = "10.1016/j.physletb.2022.137142",
    journal = "Phys. Lett. B",
    volume = "830",
    pages = "137142",
    year = "2022"
}

@article{Holdom:2022fsm,
    author = "Holdom, Bob",
    title = "{Towards rotating 2-2-holes}",
    eprint = "2208.08461",
    archivePrefix = "arXiv",
    primaryClass = "gr-qc",
    month = "8",
    year = "2022"
}

@article{DelPorro:2025wts,
    author = "Del Porro, Francesco and Pfeiffer, Jonas and Platania, Alessia and Silveravalle, Samuele",
    title = "{Charting GLOBs in Asymptotically Safe Gravity}",
    eprint = "2509.14309",
    archivePrefix = "arXiv",
    primaryClass = "gr-qc",
    month = "9",
    year = "2025"
}

@phdthesis{perkinsthesis,
      author = {Perkins, A.},
      title = {Static spherically symmetric solutions in higher derivativs gravity},
      school = {Imperial College},
      year   = {2016},  
}

@article{tHooft:1974toh,
    author = "'t Hooft, Gerard and Veltman, M. J. G.",
    title = "{One loop divergencies in the theory of gravitation}",
    journal = "Ann. Inst. H. Poincare Phys. Theor. A",
    volume = "20",
    pages = "69--94",
    year = "1974"
}

@article{Bonanno:2021zoy,
    author = "Bonanno, Alfio and Silveravalle, Samuele",
    title = "{The gravitational field of a star in quadratic gravity}",
    eprint = "2106.00558",
    archivePrefix = "arXiv",
    primaryClass = "gr-qc",
    doi = "10.1088/1475-7516/2021/08/050",
    journal = "J. Cosmol. Astropart. Phys.",
    volume = "08",
    pages = "050",
    year = "2021"
}

@book{Chandrasekhar:1985kt,
    author = "Chandrasekhar, Subrahmanyan",
    title = "{The mathematical theory of black holes}",
    isbn = "978-0-19-850370-5",
    year = "1985"
}

@article{Konoplya:2019sns,
    author = "Konoplya, R. A.",
    title = "{Shadow of a black hole surrounded by dark matter}",
    eprint = "1905.00064",
    archivePrefix = "arXiv",
    primaryClass = "gr-qc",
    doi = "10.1016/j.physletb.2019.05.043",
    journal = "Phys. Lett. B",
    volume = "795",
    pages = "1--6",
    year = "2019"
}

@article{Cardoso:2008bp,
    author = "Cardoso, Vitor and Miranda, Alex S. and Berti, Emanuele and Witek, Helvi and Zanchin, Vilson T.",
    title = "{Geodesic stability, Lyapunov exponents and quasinormal modes}",
    eprint = "0812.1806",
    archivePrefix = "arXiv",
    primaryClass = "hep-th",
    doi = "10.1103/PhysRevD.79.064016",
    journal = "Phys. Rev. D",
    volume = "79",
    number = "6",
    pages = "064016",
    year = "2009"
}

@article{Silveravalle:2022wij,
    author = "Silveravalle, Samuele and Zuccotti, Alessandro",
    title = "{Phase diagram of Einstein-Weyl gravity}",
    eprint = "2210.13877",
    archivePrefix = "arXiv",
    primaryClass = "gr-qc",
    doi = "10.1103/PhysRevD.107.064029",
    journal = "Phys. Rev. D",
    volume = "107",
    number = "6",
    pages = "064029",
    year = "2023"
}

@article{Bonanno:2022ibv,
    author = "Bonanno, Alfio and Silveravalle, Samuele and Zuccotti, Alessandro",
    title = "{Nonsymmetric wormholes and localized big rip singularities in Einstein-Weyl gravity}",
    eprint = "2204.04966",
    archivePrefix = "arXiv",
    primaryClass = "gr-qc",
    doi = "10.1103/PhysRevD.105.124059",
    journal = "Phys. Rev. D",
    volume = "105",
    number = "12",
    pages = "124059",
    year = "2022"
}

@phdthesis{Silveravalle:2023lnl,
    author = "Silveravalle, Samuele Marco",
    title = "{Isolated Objects in Quadratic Gravity: From Action Principles to Observations}",
    doi = "10.1007/978-3-031-48994-5",
    school = "Trento U.",
    year = "2024"
}

@article{Held:2022abx,
    author = "Held, Aaron and Zhang, Jun",
    title = "{Instability of spherically symmetric black holes in quadratic gravity}",
    eprint = "2209.01867",
    archivePrefix = "arXiv",
    primaryClass = "gr-qc",
    reportNumber = "Imperial/TP/2022/AH/03",
    doi = "10.1103/PhysRevD.107.064060",
    journal = "Phys. Rev. D",
    volume = "107",
    number = "6",
    pages = "064060",
    year = "2023"
}

@article{Konoplya:2011qq,
    author = "Konoplya, R. A. and Zhidenko, A.",
    title = "{Quasinormal modes of black holes: From astrophysics to string theory}",
    eprint = "1102.4014",
    archivePrefix = "arXiv",
    primaryClass = "gr-qc",
    doi = "10.1103/RevModPhys.83.793",
    journal = "Rev. Mod. Phys.",
    volume = "83",
    pages = "793--836",
    year = "2011"
}

@inproceedings{Spina:2024npx,
    author = "Spina, Andrea and Silveravalle, Samuele and Bonanno, Alfio",
    title = "{Scalar Perturbations of Regular Black Holes derived from a Non-Singular Collapse Model in Asymptotic Safety}",
    booktitle = "{17th Marcel Grossmann Meeting}: {On Recent Developments in Theoretical and Experimental General Relativity, Gravitation, and Relativistic Field Theories}",
    eprint = "2410.05936",
    archivePrefix = "arXiv",
    primaryClass = "gr-qc",
    month = "10",
    year = "2024"
}

@article{Regge:1957td,
    author = "Regge, Tullio and Wheeler, John A.",
    title = "{Stability of a Schwarzschild singularity}",
    doi = "10.1103/PhysRev.108.1063",
    journal = "Phys. Rev.",
    volume = "108",
    pages = "1063--1069",
    year = "1957"
}

@article{Zerilli:1970se,
    author = "Zerilli, Frank J.",
    title = "{Effective potential for even parity Regge-Wheeler gravitational perturbation equations}",
    doi = "10.1103/PhysRevLett.24.737",
    journal = "Phys. Rev. Lett.",
    volume = "24",
    pages = "737--738",
    year = "1970"
}

@book{Press:1992zz,
    author = "Press, William H. and Teukolsky, Saul A. and Vetterling, William T. and Flannery, Brian P.",
    title = "{Numerical Recipes in FORTRAN: The Art of Scientific Computing}",
    month = "9",
    year = "1992"
}

@book{pozrikidis2011introduction,
  title={Introduction to Theoretical and Computational Fluid Dynamics},
  author={Pozrikidis, C.},
  isbn={9780199752072},
  lccn={2011002954},
  series={EBSCO ebook academic collection},
  url={https://books.google.it/books?id=fwj1hKnLrcUC},
  year={2011},
  publisher={OUP USA}
}

@article{Berti:2007dg,
    author = "Berti, Emanuele and Cardoso, Vitor and Gonzalez, Jose A. and Sperhake, Ulrich",
    title = "{Mining information from binary black hole mergers: A Comparison of estimation methods for complex exponentials in noise}",
    eprint = "gr-qc/0701086",
    archivePrefix = "arXiv",
    doi = "10.1103/PhysRevD.75.124017",
    journal = "Phys. Rev. D",
    volume = "75",
    pages = "124017",
    year = "2007"
}

@article{Bonanno:2019rsq,
    author = "Bonanno, Alfio and Silveravalle, Samuele",
    title = "{Characterizing black hole metrics in quadratic gravity}",
    eprint = "1903.08759",
    archivePrefix = "arXiv",
    primaryClass = "gr-qc",
    doi = "10.1103/PhysRevD.99.101501",
    journal = "Phys. Rev. D",
    volume = "99",
    number = "10",
    pages = "101501",
    year = "2019"
}

@article{xhtc-9cf4,
  title = {Time evolution of black hole perturbations in quadratic gravity},
  author = "Konoplya, Roman A. and Spina, Andrea and Zhidenko, Alexander",
  journal = {Phys. Rev. D},
  pages = {--},
  year = {2025},
  month = {Jul},
  publisher = {American Physical Society},
  doi = {10.1103/xhtc-9cf4},
  url = {https://link.aps.org/doi/10.1103/xhtc-9cf4}
}

@article{Konoplya:2023fmh,
    author = "Konoplya, R. A. and Zhidenko, A.",
    title = "{Asymptotic tails of massive gravitons in light of pulsar timing array observations}",
    eprint = "2307.01110",
    archivePrefix = "arXiv",
    primaryClass = "gr-qc",
    doi = "10.1016/j.physletb.2024.138685",
    journal = "Phys. Lett. B",
    volume = "853",
    pages = "138685",
    year = "2024"
}

@article{Koyama:2000hj,
    author = "Koyama, Hiroko and Tomimatsu, Akira",
    title = "{Asymptotic power law tails of massive scalar fields in Reissner-Nordstrom background}",
    eprint = "gr-qc/0012022",
    archivePrefix = "arXiv",
    doi = "10.1103/PhysRevD.63.064032",
    journal = "Phys. Rev. D",
    volume = "63",
    pages = "064032",
    year = "2001"
}

@article{Konoplya:2006gq,
    author = "Konoplya, R. A. and Zhidenko, A. and Molina, C.",
    title = "{Late time tails of the massive vector field in a black hole background}",
    eprint = "gr-qc/0602047",
    archivePrefix = "arXiv",
    doi = "10.1103/PhysRevD.75.084004",
    journal = "Phys. Rev. D",
    volume = "75",
    pages = "084004",
    year = "2007"
}

@article{Price:1971fb,
    author = "Price, Richard H.",
    title = "{Nonspherical perturbations of relativistic gravitational collapse. 1. Scalar and gravitational perturbations}",
    doi = "10.1103/PhysRevD.5.2419",
    journal = "Phys. Rev. D",
    volume = "5",
    pages = "2419--2438",
    year = "1972"
}

@article{Price:1972pw,
    author = "Price, Richard H.",
    title = "{Nonspherical Perturbations of Relativistic Gravitational Collapse. II. Integer-Spin, Zero-Rest-Mass Fields}",
    doi = "10.1103/PhysRevD.5.2439",
    journal = "Phys. Rev. D",
    volume = "5",
    pages = "2439--2454",
    year = "1972"
}

@article{Christodoulou:1999,
  author       = {Demetrios Christodoulou},
  title        = {The instability of naked singularities in the gravitational collapse of a scalar field},
  journal      = {Annals of Mathematics},
  volume       = {149},
  number       = {1},
  pages        = {183--217},
  year         = {1999},
  doi          = {10.2307/121023},
  url          = {https://doi.org/10.2307/121023}
}

@article{Hawking:1970zqf,
    author = "Hawking, S. W. and Penrose, R.",
    title = "{The Singularities of gravitational collapse and cosmology}",
    doi = "10.1098/rspa.1970.0021",
    journal = "Proc. Roy. Soc. Lond. A",
    volume = "314",
    pages = "529--548",
    year = "1970"
}

@article{UV-grav1,
    author = "Modesto, Leonardo and Moffat, John W. and Nicolini, Piero",
    title = "{Black holes in an ultraviolet complete quantum gravity}",
    eprint = "1010.0680",
    archivePrefix = "arXiv",
    primaryClass = "gr-qc",
    doi = "https://doi.org/10.1016/j.physletb.2010.11.046",
    journal = "Phys. Lett. B.",
    volume = "695",
    pages = "297-400",
    year = "2011"
}

@article{UV-grav2,
    author = "De Lorenzo, Tommaso and Pacilio, Costantino  and Rovelli, Carlo and Speziale, Simone ",
    title = "{On the effective metric of a Planck star}",
    eprint = "1412.6015",
    archivePrefix = "arXiv",
    primaryClass = "gr-qc",
    doi = "https://doi.org/10.1007/s10714-015-1882-8",
    journal = "Gen. Relat. Gravit.",
    volume = "47",
    pages = "41",
    year = "2015"
}

@article{UV-grav3,
    author = "Perez, Alejandro",
    title = "{No firewalls in quantum gravity: the role of discreteness of quantum geometry in resolving the information loss paradox}",
    eprint = "1410.7062",
    archivePrefix = "arXiv",
    primaryClass = "gr-qc",
    doi = "https://doi.org/10.1088/0264-9381/32/8/084001",
    journal = "Class. Quantum Grav.",
    volume = "32",
    pages = "084001",
    year = "2015"
}

@article{UV-grav4,
    author = "Ashtekar, Abhay and Bojowald, Martin",
    title = "{Black hole evaporation: a paradigm}",
    eprint = "0504029",
    archivePrefix = "arXiv",
    primaryClass = "hep-th",
    doi = "https://doi.org/10.1088/0264-9381/22/16/014",
    journal = "Class. Quantum Grav.",
    volume = "22",
    pages = "3349",
    year = "2005"
}

@article{Bosma:2019aiu,
    author = "Bosma, Lando and Knorr, Benjamin and Saueressig, Frank",
    title = "{Resolving Spacetime Singularities within Asymptotic Safety}",
    eprint = "1904.04845",
    archivePrefix = "arXiv",
    primaryClass = "hep-th",
    doi = "10.1103/PhysRevLett.123.101301",
    journal = "Phys. Rev. Lett.",
    volume = "123",
    number = "10",
    pages = "101301",
    year = "2019"
}

@article{Borissova:2023kzq,
    author = "Borissova, Johanna N.",
    title = "{Suppression of spacetime singularities in quantum gravity}",
    eprint = "2309.05695",
    archivePrefix = "arXiv",
    primaryClass = "gr-qc",
    doi = "10.1088/1361-6382/ad46c0",
    journal = "Class. Quant. Grav.",
    volume = "41",
    number = "12",
    pages = "127002",
    year = "2024"
}

@article{Poisson:1988wc,
    author = "Poisson, Eric and Israel, W.",
    title = "{Structure of the Black Hole Nucleus}",
    doi = "10.1088/0264-9381/5/12/002",
    journal = "Class. Quant. Grav.",
    volume = "5",
    pages = "L201--L205",
    year = "1988"
}

@article{Harlow:2014yka,
    author = "Harlow, Daniel",
    title = "{Jerusalem Lectures on Black Holes and Quantum Information}",
    eprint = "1409.1231",
    archivePrefix = "arXiv",
    primaryClass = "hep-th",
    doi = "10.1103/RevModPhys.88.015002",
    journal = "Rev. Mod. Phys.",
    volume = "88",
    pages = "015002",
    year = "2016"
}

@inproceedings{Polchinski:2016hrw,
    author = "Polchinski, Joseph",
    title = "{The Black Hole Information Problem}",
    booktitle = "{Theoretical Advanced Study Institute in Elementary Particle Physics}: {New Frontiers in Fields and Strings}",
    eprint = "1609.04036",
    archivePrefix = "arXiv",
    primaryClass = "hep-th",
    doi = "10.1142/9789813149441_0006",
    pages = "353--397",
    year = "2017"
}
\bibliographystyle{apsrev4-1}

\end{document}